\documentclass{WileyMSP-template}
\usepackage{amsmath,amssymb,mleftright,mathtools}
\usepackage{bm}
\usepackage{microtype}
\usepackage[colorlinks,linkcolor=blue,citecolor=blue]{hyperref}
\DeclareMathAlphabet{\mathbfit}{OT1}{\familydefault}{b}{it}

\newcommand{\Hh}{\hat{H}}

\newcommand{\bt}{\bar{t}}

\newcommand{\hd}{\hat{d}}

\mathchardef\mhyphen="2D

\newcommand{\be}{\begin{equation}}
\newcommand{\ee}{\end{equation}}

\newcommand{\ii}{\mathrm{i}}

\newcommand{\mv}{\bm{v}}

\newcommand{\mh}{\bm{h}}

\newcommand{\cC}{\mathcal{C}}
\newcommand{\mPsi}{\bm{\Psi}}

\newcommand{\mPhi}{\mathbf{\Phi}}
\newcommand{\mLd}{\mathbf{\Lambda}}

\newcommand{\cG}{\mathcal{G}}

\newcommand{\mF}{\mathbf{F}}
\newcommand{\mcG}{\bm{\mathcal{G}}}

\newcommand{\mrho}{\bm{\rho}}

\newcommand{\mchi}{\bm{\mathit{\chi}}}

\usepackage{array}   % for \newcolumntype macro
\newcolumntype{L}{>{$}c<{$}} % math-mode version of "l" column type
\newcolumntype{C}{>{$}c<{$}} % math-mode version of "l" column type
\begin{document}

\pagestyle{fancy}
\rhead{\includegraphics[width=2.5cm]{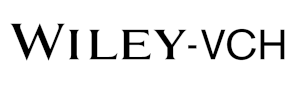}}

\title{Nonequilibrium dynamics of the Hubbard dimer}

\maketitle

% Author: Please give full first and last names for authors and include * after the name of all corresponding authors

\author{Yaroslav Pavlyukh}

\begin{affiliations}
Institute of Theoretical Physics, Faculty of Fundamental Problems of Technology,\\
Wroclaw University of Science and Technology, 50-370 Wroclaw, Poland \\
yaroslav.pavlyukh@pwr.edu.pl
\end{affiliations}

% Keywords: Please provide a minimum of three and a maximum of seven keywords, separated by commas


% Abstract should be written in the present tense and impersonal style (i.e., avoid we),
% and be at most 200 words long
\begin{abstract}
 Electron dynamics in a two-sites Hubbard model is studied using the nonequilibrium Green's function approach. The study
 is motivated by the empirical observation that a full solution of the integro-differential Kadanoff-Baym equation (KBE)
 is more stable and often accompanied by artificial damping [Marc Puig von Friesen, C. Verdozzi, and C.-O. Almbladh
   (2009)] than its time-linear reformulations relying on the generalized Kadanoff-Baym ansatz (GKBA). Additionally, for
 conserving theories, numerical simulations suggest that KBE produces natural occupations bounded by one and zero in
 agreement with the Pauli exclusion principle, whereas, in some regimes, GKBA-based theories violate this principle. As
 the first step for understanding these issues, the electron dynamics arising in the adiabatic switching scenario is
 studied.  Many-body approximations are classified according to the channel of the Bethe-Salpeter equation in which
 electronic correlations are explicitly treated. They give rise to the so-called second Born, $T$-matrix and $GW$
 approximations. In each of these cases, the model is reduced to a system of ordinary differential equations, which
 resemble equations of motion for a driven harmonic oscillator with time-dependent frequencies. A more complete
 treatment of electronic correlations is achieved by combining different correlation channels, with parquet theory
 serving as a starting point.
\end{abstract}

% Text: Please use section headings and subheadings as specified below. For
% communications, all section headings apart from Experimental Section should be removed
% Please make the first reference to a display item bold: \textbf{Figure 1} Do not
% abbreviate Figure, Equation, etc.; display items are always singular, i.e., Figure 1 and
% 2.  Equations are always singular, i.e., Equation 1 and 2, and should be inserted using
% the {equation} environment, not as graphics Please do not use footnotes in the text,
% additional information can be added to the Reference list.

%===== ===== ===== ===== ===== =====   I   ===== ===== ===== ===== ===== =====
\section{Introduction}                            
%===== ===== ===== ===== ===== ===== ===== ===== ===== ===== ===== ===== =====
Coherent electron dynamics in correlated materials attracts considerable attention due to the possibility of creating
and controlling new quantum states by time-dependent perturbations~\cite{basov_towards_2017}.  Lattice systems such as
the Hubbard model and its numerous extensions represent a versatile playground for testing many-body
theories~\cite{von_friesen_successes_2009, puig_von_friesen_kadanoff-baym_2010,karlsson_time-dependent_2011} and provide
insight into the properties of correlated materials such as cuprates, transition metal
oxides~\cite{gillmeister_ultrafast_2020} and dichalcogenides~\cite{perfetto_real-time_2022}.

Coherent dynamics in such systems can launched by periodic field driving, quenching of system parameters with electric
fields and in pump-probe experiments exploiting various combinations of phase-locked infrared (IR) and extreme
ultraviolet (XUV) pulses. Coherent dynamics can be understood on the basis of the nonequilibrium Green's function (NEGF)
approach, which can be applied perturbatively or combined with dynamical mean field theory. In a recent sequence of
works~\cite{schlunzen_achieving_2020,joost_g1-g2_2020,pavlyukh_photoinduced_2021} it has been demonstrated that
numerical NEGF approach can be significantly accelerated by the use of the so-called generalized Kadanoff-Baym ansatz
(GKBA)~\cite{lipavsky_generalized_1986} which leads to the theory formulation in the form of coupled ordinary
differential equations (GKBA+ODE).

The strength of GKBA+ODE lays in the possibility of various
extensions~\cite{pavlyukh_time-linear_2022-1,pavlyukh_time-linear_2022}, such as the inclusion of electron-phonon
interactions~\cite{karlsson_fast_2021}, multiparticle correlations~\cite{pavlyukh_photoinduced_2021}, or
transport~\cite{tuovinen_time-linear_2023}. It is also known that very systematic and balanced treatment of electronic
correlations can be achieved by working not with self-energies, but rather with vertex functions. This gives rise to the
so-called parquet method~\cite{de_dominicis_stationary_1964, pavarini_dynamical_2014}, which combines correlations in
the three channels: particle-particle ($pp$) and 2 particle-hole ($ph$, $\overline{ph}$). 

In this work, I present the unexpected finding of exact analytic solutions for the Hubbard dimer separately in all three
correlation channels within the NEGF+GKBA approximation and demonstrate the simplest possible way to combine the
channels using the so-called fluctuating-exchange approximation (FLEX).  Despite its simplicity, this model attracts
recurrent attention in the pure electronic case~\cite{romaniello_self-energy_2009,carrascal_hubbard_2015,
  mikhaylovskiy_ultrafast_2015, di_sabatino_scrutinizing_2021, joost_dynamically_2022},
linearly~\cite{berciu_exact_2007,sakkinen_many-body_2015,karlsson_fast_2021} and
quad\-ratically~\cite{kennes_transient_2017} coupled with phonons, and can be studied experimentally as ultracold atoms
in optical lattice systems~\cite{bergschneider_experimental_2019}.

The outline of the work is as follows. The GKBA+ODE approach is overviewed, presenting a uniform formulation for the
three correlation channels. Next, it is demonstrated that the respective collision integrals can be combined together
such that the double counting of Feynman diagrams can be avoided. Finally, analytic solutions for the driven Hubbard
model using all aforementioned approximations are discussed.

%===== ===== ===== ===== ===== =====   II  ===== ===== ===== ===== ===== =====
\section{Compendium of the GKBA+ODE scheme}                            
%===== ===== ===== ===== ===== ===== ===== ===== ===== ===== ===== ===== =====
Consider first a general form of the electronic Hamiltonian
\begin{align}\label{eq:H:e}
 \Hh(t)&=\sum_{ij} h_{ij}(t)\hd_{i}^\dagger \hd_{j}
+\frac12\sum_{ijmn}v_{ijmn}(t)\hd_{i}^\dagger \hd_{j}^\dagger \hd_{m} \hd_{n},
\end{align}
expressed in terms of fermionic operators $\hd^\dagger_i,\hd_j$, where $i$ may stand for
spatial degrees of freedom and spin.  In the NEGF formalism the fundamental unknowns are
the electronic lesser/greater single-particle Green's functions
\begin{align}
  G^{<}_{ij}(t,t')&= i\langle \hat{d}_j^\dagger(t')\hat{d}_i(t)\rangle,&
  G^{>}_{ij}(t,t')&=-i\langle \hat{d}_i(t)\hat{d}_j^\dagger(t')\rangle,
  \label{eq:def:G}
\end{align}
They satisfy the Kadanoff-Baym equations of motion, which are mathematically
integro-differential equations:
\begin{align}
\left[ i \partial_{t} - h_e(t) \right] G^\lessgtr(t,t')
=\left [\Sigma^\lessgtr \cdot G^A + \Sigma^R \cdot
G^\lessgtr \right]\!(t,t'),\label{eq:e:KBE}
\end{align}
where $\left[ A \cdot B\right](t,t') \equiv \int d\bt\,A(t,\bt) B(\bt,t'),$ is a real-time convolution and
$X^{R/A}(t,t')$ are the retarded/advanced functions, and $\Sigma$ is the correlation part of the self-energy. We work in
zero-temperature formalism, therefore contributions due to vertical track of the Keldysh contour are not included in
Eq.~\eqref{eq:e:KBE}. The time-local mean-field part is incorporated into the Hartree-Fock Hamiltonian
$h_{\text{HF},ij}(t)=h_{ij}(t)+\sum_{mn}w_{imnj}\rho^{<}_{nm}(t)$ with $w_{imnj}=v_{imnj}-v_{imjn}$, and we also
introduced densities according
\begin{align}
  \rho_{ij}^{\lessgtr}(t)=-i G^{\lessgtr}_{ij}(t,t).
\end{align}
By combining Eq.~\eqref{eq:e:KBE} with its adjunct and going to the equal times limit one
obtains:
\begin{align}
  \frac{d}{dt}\rho^<(t)&=-i\big[h_\text{HF}(t),\rho^<(t) \big] -\left(I(t)+I^\dagger(t)\right).
  \label{eq:eomrho:e}
\end{align}
The collision term can be expressed in terms of the two-particle Green's function (2-GF)
\begin{align}
  I_{lj}(t)&=-i\sum_{imn} v_{lnmi}(t) \cG_{imjn}(t).\label{eq:i:ee}
\end{align}
Eq.~\eqref{eq:eomrho:e} is not closed because the $\cG_{imjn}(t)$ can be expressed as a
functional of the two-times Green's function $G^{<}_{ij}(t,t')$. The complicated
time-dependence in $G^{<}_{ij}(t,t')$ is decoupled with the help of GKBA
\begin{align}
  G^{\lessgtr}(t,t')&=-G^{R}(t,t')\rho^{\lessgtr}(t')+\rho^{\lessgtr}(t)G^{A}(t,t'),\label{eq:e:gkba}
\end{align}
whereby using a simpler form of the retarded propagator:
\begin{align}
G^{R}(t,t')&=-i\theta(t-t')T\left\{e^{-\ii \int_{t'}^t d\tau\, h_\text{HF}(\tau)}\right\}.\label{eq:gr:hf}
\end{align}
Let us consider now a large class of approximations in which 2-GF is given as a solution of
the one-channel Bethe-Salpeter equation. Depending on the channel in which electronic
correlations are treated, $pp$, $ph$ or $\overline{ph}$, they are known as $T$-matrix and
$GW$ approximations. For Hubbard models, the cancellation of direct and exchange diagrams
for electrons with equal spin also allows to formulate the second Born approximation as
the common leading term of all three methods, and it can be treated on equal footing.

\begin{table}[t!]
  \caption{\label{tab:1} Definitions of electronic two-particle tensors. The vertically
    grouped indices are combined into one super-index. }
  \renewcommand{\arraystretch}{1.4}
    \begin{tabular}{LLLL}\hline
      \text{Quantity}& \text{2B and } GW & \multicolumn{1}{C}{T^{pp}} & \multicolumn{1}{C}{T^{ph}}\\\hline
      i\mchi_{\begin{subarray}{c}13\\24\end{subarray}}^{0,\lessgtr}(t,t')&
      G^{\lessgtr}_{13}(t,t')G^{\gtrless}_{42}(t',t)&
      -G^{\lessgtr}_{13}(t,t')G^{\lessgtr}_{24}(t,t')&
      -G_{13}^{\lessgtr}(t,t')G^{\gtrless}_{42}(t',t)\\  [1pt]
      \mcG_{\begin{subarray}{c}13\\24\end{subarray}}(t)&
      \mathcal{G}_{4132}(t)&
      \mathcal{G}_{1234}(t)&
      \mathcal{G}_{1432}(t)\\[1pt] 
      \mh_{\begin{subarray}{c}13\\24\end{subarray}}(t)&
      h_{13}\delta_{42}-\delta_{13}h_{42}&
      h_{13}\delta_{24}+\delta_{13}h_{24}&
      h_{13}\delta_{42}-\delta_{13}h_{42}\\ 
      \mv_{\begin{subarray}{c}13\\24\end{subarray}}(t) &
      v_{1432}(t)&
      v_{1243}(t)&
      v_{1423}(t)\\[1pt]
      \mrho^{<}_{\begin{subarray}{c}13\\24\end{subarray}}(t)&
      \rho^<_{13}(t)\rho^>_{42}(t)&
      \rho^<_{13}(t)\rho^<_{24}(t)&
      \rho^<_{13}(t)\rho^>_{42}(t)\\\hline
       \end{tabular}
\end{table}

A particularly concise formulation is achieved by introducing matrix
notations~\cite{pavlyukh_photoinduced_2021} for rank-4 tensors like 2-GF $\mcG$, Coulomb
interaction $\mv$, and the response-functions: the full $\mchi$ and the noninteracting
$\mchi^0$ ones.  The index order is different for each approximation as summarized in
Tab.~\ref{tab:1}. The one-time $\mcG(t)$ can be represented then as
% --- 
\begin{align}
\mcG(t)&=-i\int^{t}_{0}\!\!dt'\Big\{ \mchi^{>}(t,t')\mv(t')\mchi^{0,<}(t',t)
-(>\leftrightarrow<)\Big\},\label{eq:G2:e:0}
\end{align}
% ---
Here $\mchi$ is the response function describing the pure electronic screening in the
respective channel.  In Fig.~\ref{fig:GW}, the representation is illustrated for the $GW$
case in application to the Hubbard model. By differentiating Eq.~\eqref{eq:G2:e:0} with
respect to time one obtains the following equation of motion
\begin{align}
  \ii \frac{d}{dt}\mcG(t)=-\mPsi(t)+\left[\mh(t) 
   +a\mrho^\Delta(t)\mv(t)\right]\mcG(t)
  -\mcG(t) \left[\mh(t) +a\mv(t) \mrho^\Delta(t)\right],
  \label{eq:G2:X:init}
\end{align}
where $\mh$ is effective two-particle Hamiltonian. Other ingredients are defined as 
\begin{align}
  \mrho^\Delta(t)&\equiv \mrho^{>}(t)-\mrho^{<}(t),\\
  \mPsi(t)&\equiv \mrho^{>}(t)\mv(t)\mrho^{<}(t)-\mrho^{<}(t)\mv(t) \mrho^{>}(t),
  \label{rhodelta}
\end{align}
where $\mrho^{\lessgtr}$ and the lesser/greater two-particle densities.  Constant $a$ is
method-dependent and is equal to $0$, $-1$ and $1$ for second Born, $GW$ and $T$-matrix
approximations, respectively.
%===== ===== ===== ===== ===== =====  III  ===== ===== ===== ===== ===== =====
\section{Application to the Hubbard dimer}                            
%===== ===== ===== ===== ===== ===== ===== ===== ===== ===== ===== ===== =====
Specifying Eq.~\eqref{eq:H:e} to the site-spin basis $i\equiv(\bm{i},\sigma_i)$,
restricting to the nearest neighbors ($<\bm{i},\bm{j}>$) hopping
$h_{i\sigma_ij\sigma_j}=h\delta_{<\bm{i},\bm{j}>\delta_{\sigma_i\sigma_j}}$, and setting
\be
v_{\bm{i}\sigma_i \bm{j}\sigma_j \bm{k}\sigma_k \bm{l}\sigma_l}=U\delta_{\bm{ij}}\delta_{\bm{kl}}\delta_{\bm{jk}}
  \delta_{\sigma_{i}\sigma_{l}}\delta_{\sigma_{j}\sigma_{k}},\label{eq:v:def:1}
\ee
one obtains the Hubbard Hamiltonian
\be
\hat H_\text{H}=\hat T+\hat H_U=h \sum_\sigma\sum_{<\bm{i},\bm{j}>} \hd_{\bm{i}\sigma}^\dagger \hd_{\bm{j}\sigma}
+U\sum_{\bm{i}}\hat{n}_{\bm{i}\uparrow}\hat{n}_{\bm{i}\downarrow},
\ee
where $h$ is the hopping parameter, and $U$ is the on-site repulsion and
$\hat{n}_{\bm{i}\sigma}=\hd_{\bm{i}\sigma}^\dagger \hd_{\bm{i}\sigma}$. In what follows, I will focus on the case where
a lot of progress can be done analytically: the half-filled two-site Hubbard model ($N=2$ and $\sum_{\bm{i}}
\langle\hat{n}_{\bm{i}\sigma}\rangle=1$). This system has a large number of symmetries ($D_{\infty h}$ point
group). From the translational invariance follows that all physical operators are given by Toeplitz matrices. In
particular, the one-body Hamiltonian in the site basis is represented as a matrix
\be
T=\begin{pmatrix}
  0&h\\
  h&0
\end{pmatrix},\quad
\text{or for brevity}\quad T=\begin{bmatrix}
0&h
\end{bmatrix}.
\label{h0:dimer:sym}
\ee
Other symmetries (reflections) impose further restrictions on the density matrix: it is not
only Hermitian, but also symmetric and can be written in terms of just one parameter (site
basis, using shorthand notation for Toeplitz matrices)
\be
\rho_\sigma=\begin{bmatrix}1/2&a\end{bmatrix}.
\label{eq:rho:gen}
\ee
The Hartree-Fock Hamiltonian with this density reads
\be
h_\text{HF}=\begin{bmatrix}
  U/2&h
\end{bmatrix},\label{eq:h:hf}
\ee
with the eigenvalues
\be
\epsilon_i^\text{HF}=U/2\pm h.\label{eq:e:HF}
\ee
Density matrix based on the lowest energy eigenvector of Hamiltonian~\eqref{eq:h:hf} then reads
\be
\rho_\alpha^\text{HF}=\begin{bmatrix}1/2&-1/2\end{bmatrix}.
\label{eq:rho:hf}
\ee
Consider now the driving protocol in which $U=U(t)$ and $h=1$, and the density matrix possesses the full symmetry of the
system at every time instance. This discards the possibility of spontaneous dimerization. In such a scenario the density
matrix in site basis can \emph{always} be written in the form~\eqref{eq:rho:gen} with a time-dependent parameter
$a(t)$. This means that HF Hamiltonian has always the same eigenvectors, but time-dependent eigenvalues~\eqref{eq:e:HF}
(via parametric dependence of $U$ on time). Thus, HF basis is fixed. HF-Hamiltonian is time-dependent and diagonal in
this basis. Likewise, the density matrix is time-dependent and diagonal.  Since diagonal matrices commute
$[h_\text{HF}(t),\rho_\sigma(t)]=0$, in our driving protocol the density matrix is driven exclusively by the collision
term
\begin{align}
  \dot a(t)&=-2U(t)\Phi_2(t),\label{eq:rho:a}
\end{align}
where $\Phi_m(t)=-i \mathcal{G}_{11m1}(t)$ is introduced, and the initial condition according to Eq.~\eqref{eq:rho:hf}
reads
\begin{align}
  a(t_i)&=-1/2.
\end{align}
%-------------------------------------------------------------------------------------
\begin{figure}[t!]
\centering  \includegraphics[width=0.5\textwidth]{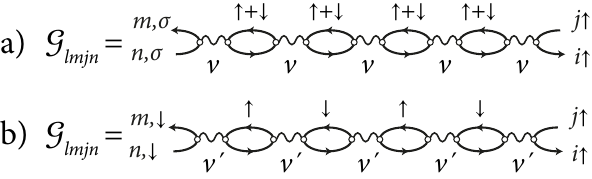}
\caption[]{Diagrammatic representation of the 2-GF ($GW$ approximation) for two forms of
  the Coulomb interaction in the Hubbard model. \label{fig:GW}}
\end{figure}
%-------------------------------------------------------------------------------------
Now the equations of motion for matrix elements of 2-GF for various approximations will be formulated and analysed.
\paragraph{Second Born approximation}
The EOM for $\Phi_2(t)$ follows from Eq.~\eqref{eq:G2:X:init}. After long but trivial calculations one obtains that 2-GF
fulfills a \emph{driven oscillator equation}
\begin{align}
  \ddot\Phi_2(t)+16\Phi_2(t)&=\dot{L}(t).\label{eq:hub2:ba}
\end{align}
with the time-dependent driving
\begin{align}
  L(t)&=U(t)\big(a(t)^3+\tfrac14a(t)\big)\equiv U(t)\lambda(t).
\end{align}
\paragraph{$T$-matrix approximations}
\begin{figure*}[th!]
  \includegraphics[width=0.24\textwidth]{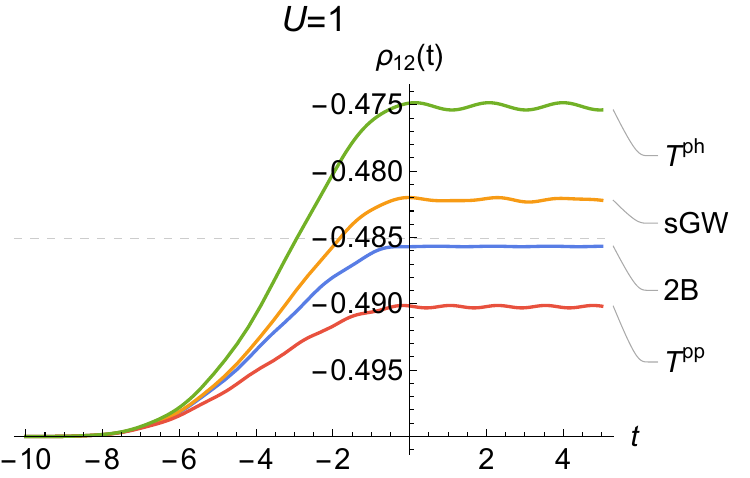}
  \hfill
  \includegraphics[width=0.24\textwidth]{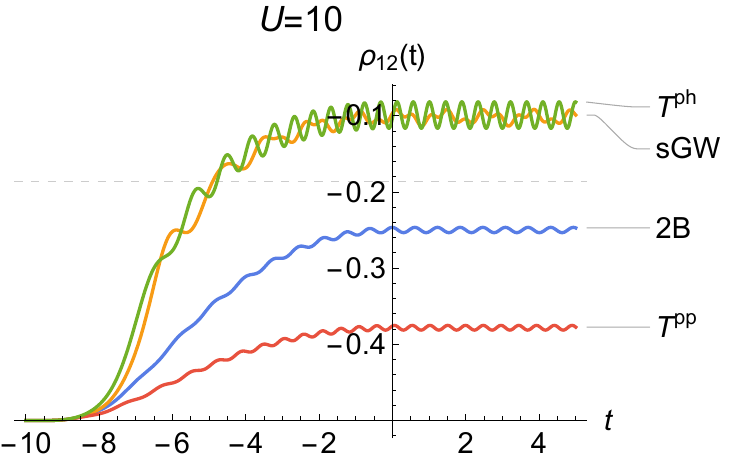}
  \hfill
  \includegraphics[width=0.24\textwidth]{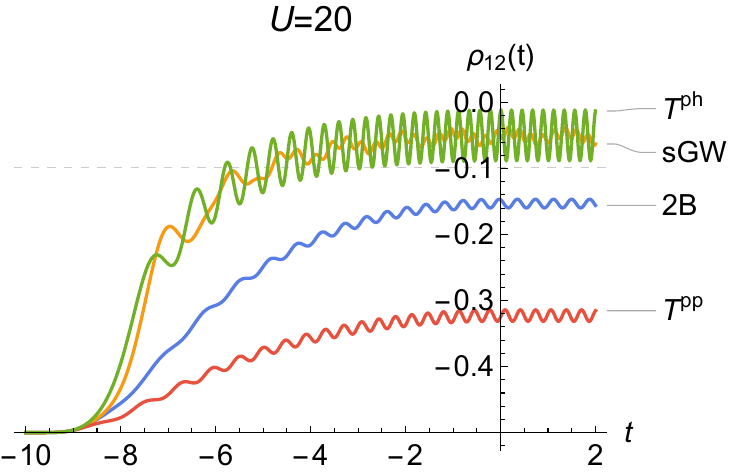}
  \hfill
  \includegraphics[width=0.24\textwidth]{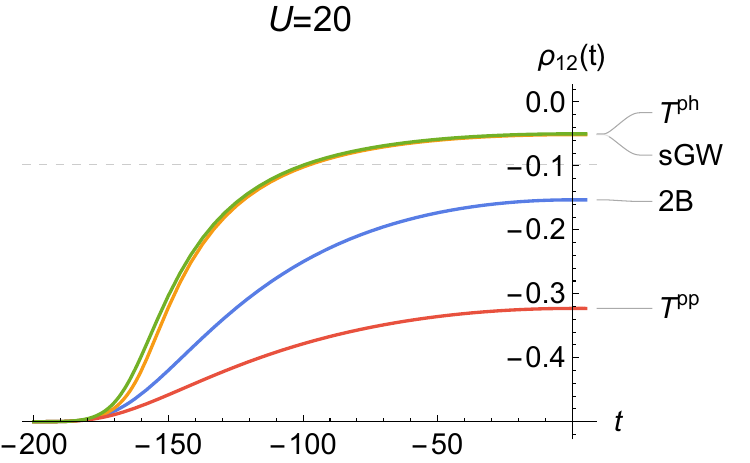}
  \hfill
\caption[]{\small The GKBA thermalization of the Hubbard dimer computed using exact
  equations for different correlated methods. It is quite surprising that $T^{ph}$ and
  s$GW$ produce very different results for small-$U$, whereas the two methods
  asymptotically converge to the same value for $U\rightarrow\infty$. Exact thermalized
  values determined by Eq.~\eqref{eq:n12:exact} are shown as dashed lines.
\label{fig:gkba:corr:analytic}} 
\end{figure*}
A marked feature of these correlated methods is that the oscillator eigenfrequency in
Eq.~\eqref{eq:hub2:ba} becomes modulated by the single-particle density. Introducing
\begin{align}
  K(t)&=U(t) a(t),
\end{align}
one obtains two coupled oscillators equations
\begin{subequations}
  \label{eq:hub2:t}
\begin{align}
  \ddot{\Phi}_2(t)+16(1\pm K(t))\Phi_2(t)&
  =\dot{L}(t)\pm 2\dot{K}(t)\Phi_1(t),\\
    \dot{\Phi}_1(t)&=-4 \Phi_2(t),
\end{align}
\end{subequations}
where the upper and lower signs correspond to the $ph$ and $pp$ channels,
respectively.  The initial conditions are
\begin{align}
  \Phi_m(t_i)&=\dot\Phi_m(t_i)=0.
\end{align}

\paragraph{$GW$ approximation} 
Eq.~\eqref{eq:v:def:1} yields the Hubbard interaction $Un_{i\uparrow}n_{i\downarrow}$ only
in conjunction with the Pauli principle. However, it is not guaranteed when working
perturbatively that the only scattering processes are those between the particles of
opposite spin, Fig.~\ref{fig:GW}(a). In this example the intermediate fermionic bubbles
may carry arbitrary spin leading to overscreening.  Same-spin scattering can be eliminated
by introducing a spin-dependent interaction~\cite{puig_von_friesen_kadanoff-baym_2010,
  thygesen_impact_2008}
\be
v'_{i\sigma_i j\sigma_j k\sigma_k l\sigma_l}=v_{i\sigma_i j\sigma_j k\sigma_k l\sigma_l}
(1-\delta_{\sigma_{i}\sigma_{j}}).
\label{eq:v:def:2}
\ee
leading to 2-GF in Fig.~\ref{fig:GW}(b). This method will be called spin-$GW$ or $sGW$. Corresponding 2-GF
equation~\eqref{eq:G2:X:init} becomes more involved due to the necessity of considering correlators with two different
spin-orders that for brevity are denoted as $a=\uparrow\downarrow\uparrow\downarrow$ and
$b=\uparrow\uparrow\uparrow\uparrow$:
\begin{subequations}
  \label{eq:hub2:sgw}
  \begin{align}
  \ddot{\Phi}^{a}_2(t)+16\Phi^{a}_2(t)&=-2\dot{K}(t)\Phi^{b}_1+16 K(t) \Phi^{b}_2 + \dot{L}(t),\\
  \ddot{\Phi}^{b}_2(t)+16\Phi^{b}_2(t)&=-2\dot{K}(t)\Phi^{a}_{1}(t)+16 K(t)\Phi^{a}_{2},\\
  \dot{\Phi}^{a}_{1}(t)&=-4 \Phi^{a}_2(t),\\
  \dot{\Phi}^{b}_{1}(t)&=-4 \Phi^{b}_2(t).
  \end{align}
\end{subequations}
Notice that the driving term is present only for the mixed spin component, which
now enters the density equation~\eqref{eq:rho:a}. 

To summarize, the EOM for the density matrix~\eqref{eq:eomrho:e} is reduced to a single differential
equation~\eqref{eq:rho:a} for a single off-diagonal element. The EOM for 2-GF~\eqref{eq:G2:X:init} takes different form
depending on the approximation: Eq.~\eqref{eq:hub2:ba} (2B), Eq.~\eqref{eq:hub2:t} ($T$-matrix), and
Eq.~\eqref{eq:hub2:sgw} ($sGW$) (see Supplemental Information for detailed derivations). Numerical solutions of these
equations for the adiabatic switching protocol
\begin{align}
  U(t)&=\begin{cases}
  U\sin^2\mleft(\tfrac{\pi}{2}\tfrac{(t+\tau)}{\tau}\mright)&\tau<t<0,\\
  U&t\ge0;
  \end{cases}
\end{align}
and different final Hubbard-$U$ and switching times ($t_i=\tau$) are shown in Fig.~\ref{fig:gkba:corr:analytic}. There
are two observations in comparison with earlier works using the full KBE propagation. First of all, a crucial difference
between the GKBA method and the full solution of KBE is that the former does not lead to the artificial damping observed
in the paper of Marc Puig von Friesen, C. Verdozzi, and C.-O. Almbladh~\cite{von_friesen_successes_2009}. Besides the
numerical evidences, this observation is supported by the structure of the obtained equations, which have a driven
oscillator form without a damping term. The damping was discussed in another papers by the same
authors~\cite{puig_von_friesen_kadanoff-baym_2010, friesen_artificial_2010}, and a plausible explanation is
that it results from self-consistent treatments, leading to an infinite number of poles in the electron Green’s
function. In contrast, an exact solution should only have a finite number of poles for finite systems. Another
interesting effect is the lack multiple steady
states~\cite{von_friesen_successes_2009,puig_von_friesen_kadanoff-baym_2010}. This is also an artefact of exact
calculation not observed in the GKBA scheme. As comparison of Fig.~\ref{fig:gkba:corr:analytic}(c) and (d) shows, fast
switching of the interaction leads to oscillations, but there is never a transition to a state different from the one
obtained by slow adiabatic switching. This holds true even when artificial damping term is added. However, it is not
possible to exclude that under some combination of parameters an artificial steady state can be reached.

\subsection{Asymptotic analysis}
Starting from the zero-temperature adiabatic assumption the correlated density matrix of the ground state of the system
can be determined analytically for 2B and $T$-matrix methods. To this end equations of motion are written in the form common
to all methods, the $\ddot{\Phi}_2(t)$ is neglected assuming infinitesimally slow switching, parametrix dependence of
physical quantities on $U$ is introduced, i.e., $a(U)\equiv a(U(t))$, and $f(u)\equiv \Phi_1(U(t))$. We obtain
\begin{subequations}
\begin{align}
  2D_U[a(U)]&=U D_U[f(U)],\\
  -4 D_U[f(U)] (1-a(U) \eta U)&=D_U[u\lambda(a(U))]-2\eta f(U)D_U[U a(U)],\\
  a(0)&=-1/2, f(0)=0.
\end{align}
\end{subequations}
where $\eta=1, -1, 0$ for $T^{pp}$, $T^{ph}$, 2B approximations, respectively. $D_U$ denotes the derivative with respect
to $U$ and $\lambda(a)=a^3+\frac14a$. Surprisingly, these equations can be analytically integrated (see Supplemental
Information for detailed derivations) leading to
\begin{subequations}
  \label{aU:implicit}
\begin{align}
  a^2\left[16 + 16 \left(1 + 4 a^2\right) + U^2 \left(1 + 4 a^2\right)^2\right]&=12,&& \text{2B};\label{eq:a:2ba}\\
  a^2 (U a+2)^2&=U a+1,&& T^{ph};\label{eq:a:tph}\\
  a \left(a (U a -2)^2+U\right)&=1,&& T^{pp}.\label{eq:a:tpp}
\end{align}
\end{subequations}
The derivation for the spin $GW$ method proceeds along the same line, except there are more equations to solve. As the
consequence, it is not possible to integrate all of them. However, one can reduce (Supplemental Information) them to a
single nonlinear ordinary differential equation
\begin{align}
  p(a, U) a'(U)^2 + q(a, U) a'(U) + r(a, U)&=0,\label{eq:aU:sGW:ODE}
\end{align}
where
\begin{align*}
  p(a, U) &= 3 U^2 a^2 \left(36 U^4 a^4+3 \left(U^2-48\right) U^2 a^2-U^4+8  U^2+192\right)-\left(U^2+16\right)^2,\\
  q(a, U) &= 2 U a \left(9 U^4 a^4-3 U^2 \left(U^2-4\right) a^2-(U^2+16)\right),\\
  r(a, U)&= U^2 a^2 \left(3 U^2 a^2 \left(3 a^2-1\right)-1\right).
\end{align*}
While we were not able to solve Eq.~\eqref{eq:aU:sGW:ODE} analytically, it is possible to perform series expansions and
to compare with the exact solution~\cite{stefanucci_nonequilibrium_2013} that has much simpler form
\be
a(U)=-\frac{2}{\sqrt{U^2+16}}.\label{eq:n12:exact}
\ee
From the implicit equations~\eqref{aU:implicit}, differential equation~\eqref{eq:aU:sGW:ODE} and algebraic
equation~\eqref{eq:n12:exact} we obtain the series expansions for $U\rightarrow 0$
\begin{subequations}
  \label{aU:U0}
\begin{align}
  a&=-\frac{1}{2}+\frac{U^2}{64}-\frac{3 U^4}{2048}+\mathcal{O}\mleft(U^6\mright),&\text{2B};\\
  a&=-\frac{1}{2}+\frac{U^2}{64}+\frac{9 U^4}{4096}+\mathcal{O}\mleft(U^6\mright)& sGW;\\
  a&=-\frac{1}{2}+\frac{U^2}{64}+\frac{U^3}{128}+\frac{9 U^4}{4096}+\mathcal{O}\mleft(U^6\mright),&T^{ph};\\
  a&=-\frac{1}{2}+\frac{U^2}{64}-\frac{U^3}{128}+\frac{9 U^4}{4096}+\mathcal{O}\mleft(U^6\mright),&T^{pp};\\
  a&=-\frac{1}{2}+\frac{U^2}{64}-\frac{3 U^4}{4096}+\mathcal{O}\mleft(U^6\mright),&\text{exact}.
\end{align}
\end{subequations}
All of them are exact of to the second order. This is expected because the second order diagrams are fully taken into
account by all the methods. $T$-matrix methods contain terms of the thirds order corresponding to the ladder diagrams
with three interaction lines. Exact result does not have these terms. Thus, these diagrams must be compensated by other
third order diagrams not accessible with our method. The spin $GW$ method contains only the even order terms. This is
expected from the diagrammatic construction illustrated in Fig.~\ref{fig:GW}. 2B method also contains only even order
terms. Terms of fourth order and higher arise from the self-consistent solution of the Dyson equation.

\begin{figure}[t!]
  {
  \centering
  \hfill
  \includegraphics[width=0.38\textwidth]{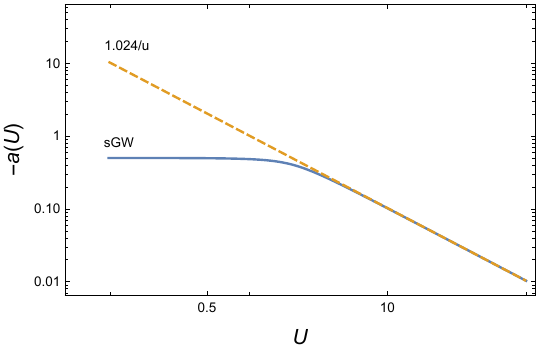}
  \hfill
  \includegraphics[width=0.52\textwidth]{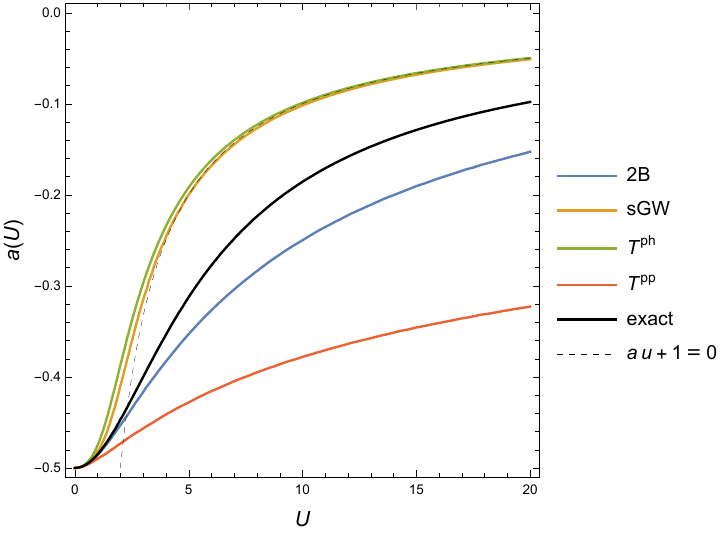}
  \hfill}
\caption[]{Asymtpotics of the off-diagonal density matrix element $a(U)$ for the spin-$GW$ method (left) and comparison of
  $a(U)$ for different methods (right). \label{fig:aU:contours}}
\end{figure}

For large values of $U$ the off-diagonal density approaches zero as
\begin{subequations}
  \label{aU:Uinf}
\begin{align}
  a&=-\frac{2 \sqrt{3}}{U}+\mathcal{O}\mleft({U}^{-3}\mright),&\text{2B};\\
  a&=-\frac{1.024}{U}+\mathcal{O}\mleft({U}^{-3}\mright),&sGW;\\
  a&=-\frac{1}{U}+\mathcal{O}\mleft({U}^{-3}\mright),&T^{ph};\\
  a&=-\frac{1}{U^{1/3}}+ \frac{1}{U}+\mathcal{O}\mleft({U}^{-3}\mright),&T^{pp};\\
  a&=-\frac{2}{U}+\mathcal{O}\mleft({U}^{-3}\mright),&\text{exact}.
\end{align}
\end{subequations}
It is quite unexpected that $T^{pp}$ approximation fails to provide even the $-1/U$ asymptotic dependence of the exact
solution. It should also be noticed that spin-$GW$ is numerically very close to the $T^{ph}$ method, however, the
asymptotic coefficient (determined numerically) is slightly different [$-1.024$, see
  Fig.~\ref{fig:aU:contours}(left)]. This indicates that physically the two methods are physically different, and the
observed similarity is rather a coincidence.

%-------------------------------------------------------------------------------------
%-------------------------------------------------------------------------------------

Analysing further the $a(U)$ dependence based on Eqs.~\eqref{aU:implicit} allows to conclude that physical (fulfilling
the initial condition) branch of the considered theories always satisfies $-1/2<a(U)\le0$ in equilibrium. This indicates
that natural occupations $n_i(U)$ are always in the range $0\le n_i\le 1$, as stipulated by the Pauli exclusion
principle. This is a non-trivial finding as apart from the Hartree-Fock approximation there is no proof that the natural
occupations should adhere to physical limits in correlated theories treated within GKBA. In fact, there are
electronic~\cite{joost_dynamically_2022} and electron-phonon~\cite{pavlyukh_time-linear_2022} systems, where under some
conditions violations of these limits occur.

It is also interesting to note that the structure of time-dependent equations~\eqref{eq:hub2:t} is very similar to the
structure of electron self-energies derived for the same system in the ground state~\cite{romaniello_beyond_2012}. For
instance for $T^{ph}$ and $T^{pp}$, the self-energy poles (see Eqs.(53, 43) therein) are expressed in terms of the
effective Hamiltonian that reduces to the double effective frequency in Eq.~\eqref{eq:hub2:t} ($\Omega^2=1\pm aU $) when
Hartree-Fock $a=-1/2$ is used therein. It also explains why the $T^{ph}$ theory of Ref.~\cite{romaniello_beyond_2012} is
unstable when $U$ approaches 2, and why such an instability is absent in our case. As can be see from
Fig.~\ref{fig:aU:contours}(right), in our approach $a$ depends on $U$ and therefore the effective frequency $1+aU$
(dashed line) never becomes equal to zero.
%===== ===== ===== ===== ===== =====  III  ===== ===== ===== ===== ===== =====
\section{Combining channels}                            
%===== ===== ===== ===== ===== ===== ===== ===== ===== ===== ===== ===== =====

\begin{figure}
 \centering   \includegraphics[width=0.45\textwidth]{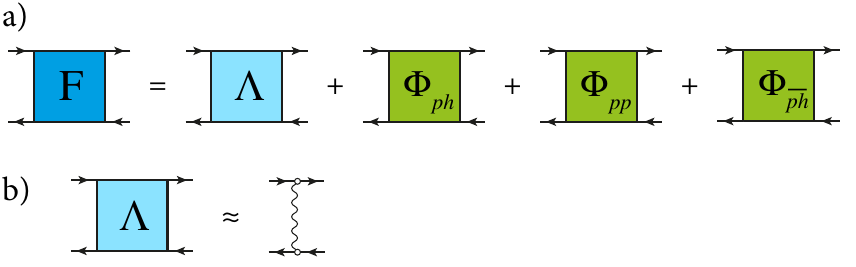}
\caption[]{\small (a) Vertex function as a sum of irreducible and two-particle reducible
  correlators, Eq.~\eqref{eq:F:def}. (b) Approximation of the lowest-order irreducible
  vertex by the direct Coulomb interaction (exchange is not included).
\label{fig:vertex}}
\end{figure}

Numerical examples shown above indicate that $sGW$, $T^{pp}$ and $T^{ph}$ approximations, despite containing infinite
sequence of Feynman diagrams via the solution of the Bethe-Salpeter equation in the respective channels, are describing
physically very different scenarios. Can one exploit the advantages of each approximation? In the following, I derive
such an approximation starting from the exact parquet equations~\cite{pavarini_dynamical_2014}. This derivation is
complementary to the self-energy based derivation in Ref.~\cite{pavlyukh_interacting_2022}. The motivation for this
starting point is two-fold: on one side this derivation establishes that there is no double counting of Feynman
diagrams. On the other side, it demonstrates difficulties of going beyond. They are associated with multiple
time-arguments of constituent vertex functions~\cite{yan_single-impurity_2022}. In particular, working in one channel
allows one to express the one-time 2-GF in terms of two-times response functions, see for instance Eq.~(28) of
Ref.~\cite{pavlyukh_photoinduced_2021}, and to close the equation of motion for it. Combining the channels requires in
general to deal with 4-times quantities for which GKBA is not known.

Consider the full (reducible) vertex $\mF$, which is just 2-GF with amputated fermionic
lines, i.\,e., $i\mcG=\mchi^0\mF\mchi^0$. This equation is written schematically as it
does not reflect the time arguments of the ingredients. Reducing $\mcG$ to two-times form
is the main goal here.  Quite generally, the \emph{vertex} function $\mF$ can be written
as a sum of the \emph{fully irreducible vertex} $\mLd$ and reducible vertices $\mPhi_i$
\begin{align}
\mF&=\mLd+\sum\nolimits_i \mPhi_i,\label{eq:F:def}
\end{align}
where $i=pp,\,ph,\,\overline{ph}$ denotes the channels, in which $ \mPhi_i$ are
\emph{two-particle reducible}. Eq.~\eqref{eq:F:def} is, therefore, only a topological
statement (Fig.~\ref{fig:vertex}). It is important because it rules out any double
counting of the resulting Feynman diagrams.

\begin{figure}
 \centering   \includegraphics[width=0.45\textwidth]{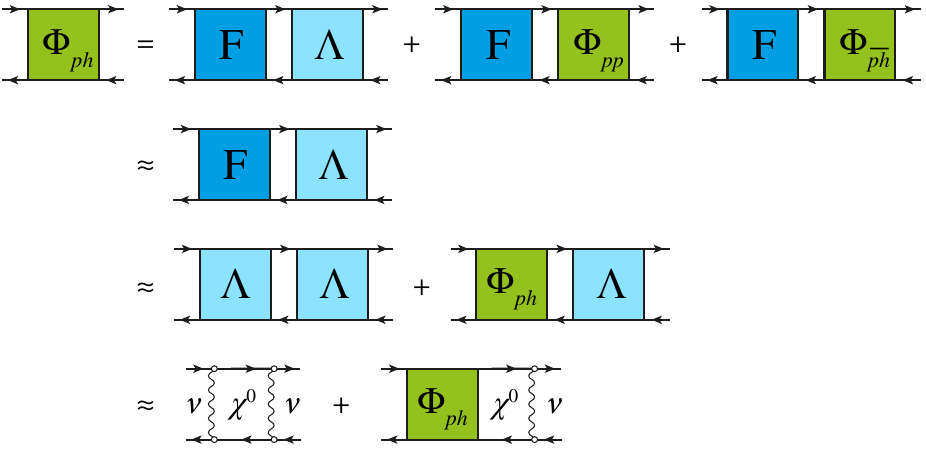}
\caption[]{\small Derivation of the Bethe-Salpeter equation~\eqref{eq:F:eom} for the
  two-particle $ph$ reducible correlator. $\overline{ph}$ and $pp$ channels are treated in
  the same way. Notice that the first approximation means setting the irreducible vertex
  in the $ph$ channel to the fully irreducible vertex (and similarly in other
  channels). This is exactly what defines the FLEX
  approximation~\cite{rohringer_local_2012}.
\label{fig:Phi:ph}}
\end{figure}

The reducible vertices $\mPhi_i$ fulfill a set of interrelated equations (see first line of Fig.~\ref{fig:Phi:ph} for
the exact equation in the $ph$ channel). Two approximations can be introduced allowing to transit from four to two
time-arguments: (i) The leading term for each equation is written as $(\mF\mchi^0\mLd)_i$ (second line of
Fig.~\ref{fig:Phi:ph}) in all channels. Here, the subscript $i$ indicates that quantities in the brackets are given in
the index order pertinent to the channel, other terms are discarded; (ii) Approximating therein
\begin{align}
 \mF&\approx \mLd + \mPhi_i,\label{eq:F0:def}
\end{align}
(third line) and using $\mLd\approx \mv$ [Fig.~\ref{fig:vertex}(b)] leads to the
Bethe-Salpeter equation for the two-particle reducible functions (forth line of
Fig.~\ref{fig:Phi:ph})
\begin{align}
  \mPhi_i(z,z')&\approx\mv(z)\mchi_i^0(z,z')\mv(z')
  + \int_{\cC} d\bar{z}\,\mPhi_i(z,\bar{z})\mchi_i^0(\bar{z},z')\mv(z'),\label{eq:F:eom}
\end{align}
where $z$, $z'$ are times on the Keldysh contour $\cC$. $\mchi_i^0$ is written explicitly
for each channel in Tab.~\ref{tab:1}, where $GW$ approximation corresponds to the
treatment of $\overline{ph}$ channel, $T^{pp}$ approximation\,---\,$pp$ channel, and
$T^{ph}$\,---\,$ph$ channel. Notice that while exact Bethe-Salpeter equations are
formulated for 4-times correlators, the final approximated equation is closed for
$\mPhi_i$ dependent on two-times.

Decorating $\mPhi_i$ with $\mchi_i^0$ on both sides of Eq.~\eqref{eq:F:eom} one obtains the RPA equations for the full
response functions $\mchi_i(z,z')\mv(z')=[\mchi_i^0\cdot(\mv+\mPhi_i)](z,z')$:
\begin{align}
  \mchi_i(z,z')&=\mchi_i^0(z,z')
  + \int_{\cC} d\bar{z}\,\mchi_i(z,\bar{z})\mv(\bar{z})\mchi_i^0(\bar{z},z').\label{eq:rpa}
\end{align}

Let us now decorate the vertex functions in Eq.~\eqref{eq:F0:def} with four fermionic
lines and go to the equal time-limit ($t$) for the external time-arguments
\begin{align}
  i\mcG_i(t)=[\mchi^0_i\cdot\mv\cdot\mchi^0_i+\mchi^0_i\cdot\mPhi_i\cdot\mchi^0_i](t)=[\mchi_i\cdot\mv\cdot\mchi^0_i](t).
\end{align}
In this way Eq.~\eqref{eq:G2:e:0} was re-derived for each channel. However, one can do better by using the \emph{full}
vertex as in Eq.~\eqref{eq:F:def} instead of the \emph{partial} vertex as in Eq.~\eqref{eq:F0:def}. To this end, let us
decorate Eq.~\eqref{eq:F:def} ($\mLd\approx \mv$) with four fermionic lines with equal external time-arguments ($t$). It
follows then
\begin{align}
  \cG(t)&=-2\cG_\text{2B}(t)+\sum\nolimits_i\cG_i(t).\label{eq:i}
\end{align}
Notice that the matrix notations (Tab.~\ref{tab:1}) are not used here because each approximation has own order of
indices. Instead, $\cG$s are interpreted here as usual rank-four tensors, like in Eq.~\eqref{eq:i:ee}. A crucial point
of this derivation is to realize that
\begin{align}
  i\mcG_\text{2B}(t)&=[\mchi_i^0\cdot\mv\cdot\mchi_i^0](t)
\end{align}
is independent on the channel $i$ and is given simply by the second Born approximation for
2-GF. This can be verified by using definitions in Tab.~\ref{tab:1} ( provided that lhs of
this equation is written with index order pertinent to the rhs). Therefore
Eq.~\eqref{eq:i} can be written explicitly for the Hubbard model with spin-dependent
interactions~\eqref{eq:v:def:2} as
\begin{align}
  \cG_\text{FLEX}(t)&=-2\cG_\text{2B}(t)+\cG_{sGW}(t)+\cG_{T^{pp}}(t)+\cG_{T^{ph}}(t).\label{eq:flex}
\end{align}
 The resulting approximation is denoted as FLEX\,---the fluctuating-exchange approximation\,---\,which seems to be the
 common term~\cite{bickers_conserving_1989,drchal_dynamical_2005,joost_g1-g2_2020}. It may be viewed as a first-order
 element in a hierarchy of successive approximations to the full parquet
 solution~\cite{bickers_conserving_1991,rohringer_local_2012}.

\begin{figure}[t!]
  \includegraphics[width=0.31\textwidth]{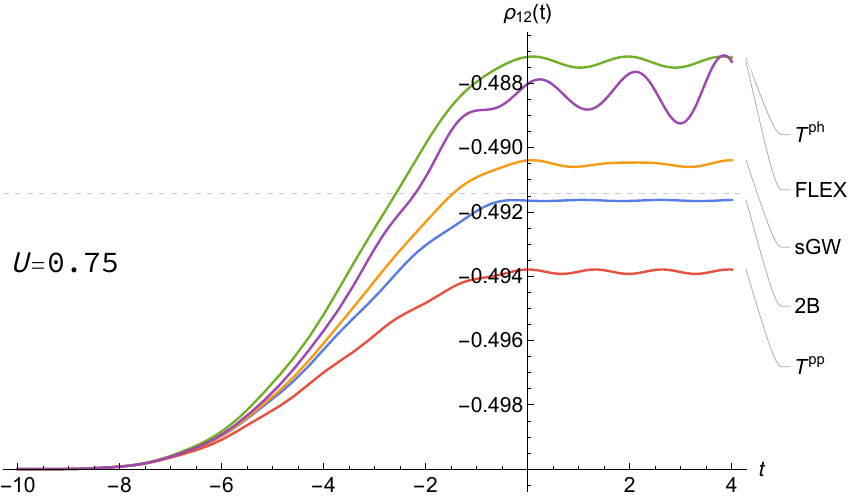}
  \hfill
  \includegraphics[width=0.31\textwidth]{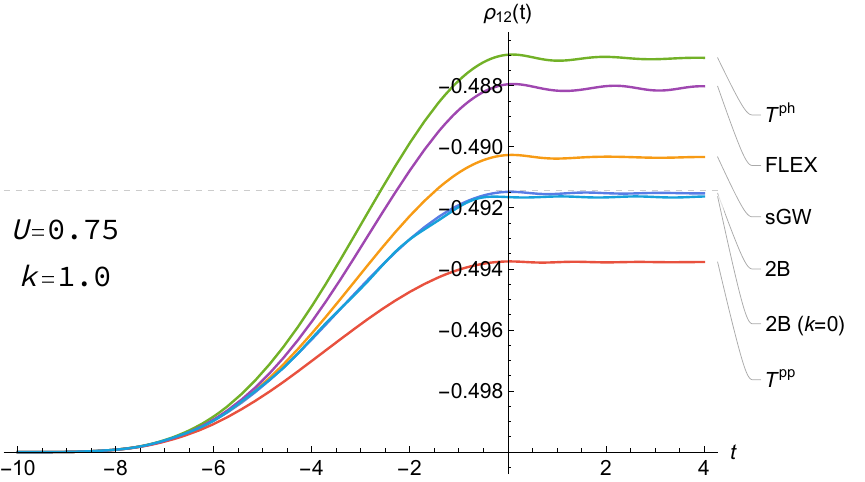}
  \hfill
  \includegraphics[width=0.31\textwidth]{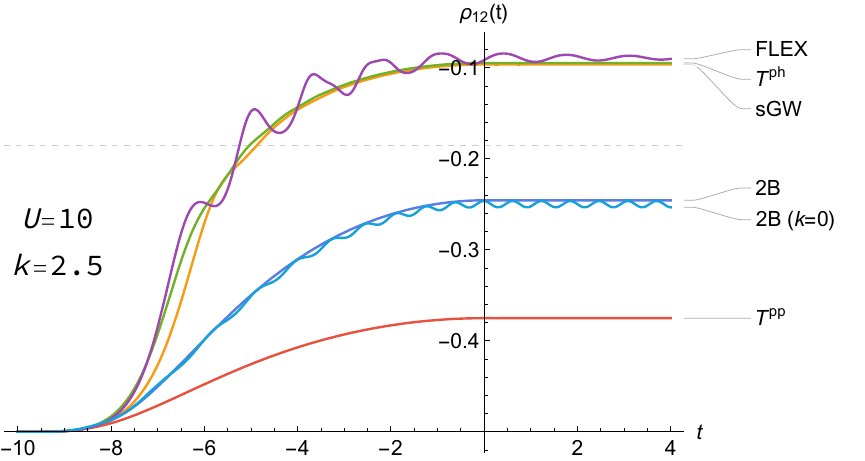}
\caption[]{\small The GKBA thermalization of the Hubbard dimer computed using exact
  equations for different correlated methods. Dashed horizontal line depicts exact
  thermalized value of $a(t)$ given by Eq.~\eqref{eq:n12:exact}.
\label{fig:gkba:flex}}
\end{figure}

Thermalization is possible only for small values of $U$ as depicted in Fig.~\ref{fig:gkba:flex} (left). There is,
however, a simple procedure that stabilizes the GKBA+ODE methods. It was already mentioned that EOMs for the matrix
elements of 2-GF have a driven harmonic oscillator form, viz. Eq.~(\ref{eq:hub2:ba}, \ref{eq:hub2:t},
\ref{eq:hub2:sgw}). By introducing a ``velocity damping'' with coefficient $k$ (see Supporting Information for the exact
form of equations) converged solutions can be obtained even for strongly correlated cases, Fig.~\ref{fig:gkba:flex}
(right). If for a given method the adiabatic switching procedure can be performed, then the final density matrix of the
system is characterised by the $a(U)$ value independent of a small damping $k$. The term “small” stipulates that the
magnitude of damping should not exceed other energy scales in the system (see Supporting Information for the
investigation of the role of $k$). While the ``damping'' approach work excellently for the Hubbard dimer, further work
is needed in order to generalize this approach to more complicated systems.

Finally, we remark that our approach is different from the dynamically-screened-ladder
approximation in Ref.~\cite{joost_g1-g2_2020} in that four ingredient 2-GFs in
Eq.~\eqref{eq:flex} are independently propagated.
\section{Conclusions and outlook}
Exact many-body electron dynamics in a finite basis representation is described by the linear differential equation for
the wave-function. Reformulations of this dynamics on the language of reduced quantities, such as nonequilibrium Green's
functions, inevitably lead to the equations of motion with much more complicated mathematical structure. One of the
first attempts to analyze the integro-differential Kadanoff-Baym equations in this perspective have been made by Marc
Puig von Friesen, C. Verdozzi, and C.-O. Almbladh~\cite{puig_von_friesen_kadanoff-baym_2010}. They have made a number of
interesting observations concerning non-unique steady states and the role of the correlation-induced damping. These
findings are revisited here in view of the rapid development of the time-linear methods that can be formulated as a
system of coupled ordinary differential equations.  In this work, the first steps are taken in the investigation of the
nonlinear form of the GKBA+ODE scheme by applying it to the two-sites Hubbard model. This involves deriving
oscillator-like equations for the components of the two-particle GF and analyzing the adiabatic switching scenario and
the steady-state limit. This has led to the observation of similarities between equations in different channels,
systematic under- and over-estimations of the correlational effects pertinent to different schemes and very close but
distinct asymptotic limits for the $T^{ph}$ and $sGW$ methods. It was also possible to prove analytically that second
Born, $T$-matrix and spin-$GW$ approximations never violate the Pauli exclusion principle in the steady state limit.
Nonetheless, such violations may take place under strongly nonequilibrium conditions. Further investigations of these
effects are needed. The structure of the GKBE+ODE equations in the dimer case hints that this is a challenging
mathematical problem.

\section*{Supporting Information}
Supporting Information is available from the Wiley Online Library or from the author. 

\section*{Acknowledgements}
I would like to thank Enrico Perfetto and Gianluca Stefanucci for insightful discussions, and Hugo U.R. Strand, Claudio
Verdozzi, Michael Bonitz for the excellent organization of the PNGF8 conference. This research was part of project
no. 2021/43/P/ST3/03293 cofunded by the National Science Centre and the European Union’s Horizon 2020 research and
innovation programme under the Marie Sklodowska-Curie grant agreement no. 945339.

% References
\medskip

% Use the following code if you wish to generate your bibliography with BibTeX;
% replace the string "MSP-template" below with the name(s) of
% the BibTeX data base(s) you want to use.
% The resulting bibliography-output (the content of the .bbl file)
% must be pasted back into this file before submission.
% Please also include your BibTeX data base file(s) in your submission
% so that we can re-run BibTeX if necessary.
%
%\bibliographystyle{MSP}
%\bibliography{MyLibrary}

\begin{thebibliography}{10}
\providecommand{\url}[1]{\texttt{#1}}
\providecommand{\urlprefix}{URL }

\bibitem{basov_towards_2017}
D.~N. Basov, R.~D. Averitt, D.~Hsieh,
\newblock \emph{Nat. Mater.} \textbf{2017}, \emph{16}, 11 1077.

\bibitem{von_friesen_successes_2009}
M.~P. von Friesen, C.~Verdozzi, C.-O. Almbladh,
\newblock \emph{Phys. Rev. Lett.} \textbf{2009}, \emph{103}, 17 176404.

\bibitem{puig_von_friesen_kadanoff-baym_2010}
M.~Puig~von Friesen, C.~Verdozzi, C.-O. Almbladh,
\newblock \emph{Phys. Rev. B} \textbf{2010}, \emph{82}, 15 155108.

\bibitem{karlsson_time-dependent_2011}
D.~Karlsson, A.~Privitera, C.~Verdozzi,
\newblock \emph{Phys. Rev. Lett.} \textbf{2011}, \emph{106}, 11 116401.

\bibitem{gillmeister_ultrafast_2020}
K.~Gillmeister, D.~Gole\v{z}, C.-T. Chiang, N.~Bittner, Y.~Pavlyukh,
  J.~Berakdar, P.~Werner, W.~Widdra,
\newblock \emph{Nat. Commun.} \textbf{2020}, \emph{11}, 1 4095.

\bibitem{perfetto_real-time_2022}
E.~Perfetto, Y.~Pavlyukh, G.~Stefanucci,
\newblock \emph{Phys. Rev. Lett.} \textbf{2022}, \emph{128}, 1 016801.

\bibitem{schlunzen_achieving_2020}
N.~Schl\"{u}nzen, J.-P. Joost, M.~Bonitz,
\newblock \emph{Phys. Rev. Lett.} \textbf{2020}, \emph{124}, 7 076601.

\bibitem{joost_g1-g2_2020}
J.-P. Joost, N.~Schl\"{u}nzen, M.~Bonitz,
\newblock \emph{Phys. Rev. B} \textbf{2020}, \emph{101}, 24 245101.

\bibitem{pavlyukh_photoinduced_2021}
Y.~Pavlyukh, E.~Perfetto, G.~Stefanucci,
\newblock \emph{Phys. Rev. B} \textbf{2021}, \emph{104}, 3 035124.

\bibitem{lipavsky_generalized_1986}
P.~Lipavský, V.~\v{S}pi\v{c}ka, B.~Velický,
\newblock \emph{Phys. Rev. B} \textbf{1986}, \emph{34}, 10 6933.

\bibitem{pavlyukh_time-linear_2022-1}
Y.~Pavlyukh, E.~Perfetto, D.~Karlsson, R.~van Leeuwen, G.~Stefanucci,
\newblock \emph{Phys. Rev. B} \textbf{2022}, \emph{105}, 12 125134.

\bibitem{pavlyukh_time-linear_2022}
Y.~Pavlyukh, E.~Perfetto, D.~Karlsson, R.~van Leeuwen, G.~Stefanucci,
\newblock \emph{Phys. Rev. B} \textbf{2022}, \emph{105}, 12 125135.

\bibitem{karlsson_fast_2021}
D.~Karlsson, R.~van Leeuwen, Y.~Pavlyukh, E.~Perfetto, G.~Stefanucci,
\newblock \emph{Phys. Rev. Lett.} \textbf{2021}, \emph{127}, 3 036402.

\bibitem{tuovinen_time-linear_2023}
R.~Tuovinen, Y.~Pavlyukh, E.~Perfetto, G.~Stefanucci,
\newblock \emph{Phys. Rev. Lett.} \textbf{2023}, \emph{130}, 24 246301.

\bibitem{de_dominicis_stationary_1964}
C.~De~Dominicis, P.~C. Martin,
\newblock \emph{J. Math. Phys.} \textbf{1964}, \emph{5}, 1 14.

\bibitem{pavarini_dynamical_2014}
K.~Held,
\newblock In E.~Pavarini, E.~Koch, D.~Vollhardt, A.~I. Lichtenstein, editors,
  \emph{{DMFT} at 25: infinite dimensions: lecture notes of the {Autumn}
  {School} on {Correlated} {Electrons} 2014}. Forschungszentrum J\"{u}lich,
  Zentralbibliothek, Verl, J\"{u}lich, \textbf{2014}.

\bibitem{romaniello_self-energy_2009}
P.~Romaniello, S.~Guyot, L.~Reining,
\newblock \emph{J. Chem. Phys.} \textbf{2009}, \emph{131}, 15 154111.

\bibitem{carrascal_hubbard_2015}
D.~J. Carrascal, J.~Ferrer, J.~C. Smith, K.~Burke,
\newblock \emph{J. Phys. Condens. Matter} \textbf{2015}, \emph{27}, 39 393001.

\bibitem{mikhaylovskiy_ultrafast_2015}
R.~Mikhaylovskiy, E.~Hendry, A.~Secchi, J.~Mentink, M.~Eckstein, A.~Wu,
  R.~Pisarev, V.~Kruglyak, M.~Katsnelson, T.~Rasing, A.~Kimel,
\newblock \emph{Nat. Commun.} \textbf{2015}, \emph{6}, 1 8190.

\bibitem{di_sabatino_scrutinizing_2021}
S.~Di~Sabatino, P.-F. Loos, P.~Romaniello,
\newblock \emph{Frontiers in Chemistry} \textbf{2021}, \emph{9} 751054.

\bibitem{joost_dynamically_2022}
J.-P. Joost, N.~Schl\"{u}nzen, H.~Ohldag, M.~Bonitz, F.~Lackner,
  I.~B\v{r}ezinov\'{a},
\newblock \emph{Phys. Rev. B} \textbf{2022}, \emph{105}, 16 165155.

\bibitem{berciu_exact_2007}
M.~Berciu,
\newblock \emph{Phys. Rev. B} \textbf{2007}, \emph{75}, 8 081101(R).

\bibitem{sakkinen_many-body_2015}
N.~S\"{a}kkinen, Y.~Peng, H.~Appel, R.~van Leeuwen,
\newblock \emph{J. Chem. Phys.} \textbf{2015}, \emph{143}, 23 234102.

\bibitem{kennes_transient_2017}
D.~M. Kennes, E.~Y. Wilner, D.~R. Reichman, A.~J. Millis,
\newblock \emph{Nature Phys.} \textbf{2017}, \emph{13}, 5 479.

\bibitem{bergschneider_experimental_2019}
A.~Bergschneider, V.~M. Klinkhamer, J.~H. Becher, R.~Klemt, L.~Palm,
  G.~Z\"{u}rn, S.~Jochim, P.~M. Preiss,
\newblock \emph{Nature Phys.} \textbf{2019}, \emph{15}, 7 640.

\bibitem{thygesen_impact_2008}
K.~S. Thygesen,
\newblock \emph{Phys. Rev. Lett.} \textbf{2008}, \emph{100}, 16 166804.

\bibitem{friesen_artificial_2010}
M.~P.~v. Friesen, C.~Verdozzi, C.-O. Almbladh,
\newblock \emph{J. Phys. Conf. Ser.} \textbf{2010}, \emph{220} 012016.

\bibitem{stefanucci_nonequilibrium_2013}
G.~Stefanucci, R.~van Leeuwen,
\newblock \emph{Nonequilibrium {Many}-{Body} {Theory} of {Quantum} {Systems}:
  {A} {Modern} {Introduction}},
\newblock Cambridge University Press, Cambridge, \textbf{2013}.

\bibitem{romaniello_beyond_2012}
P.~Romaniello, F.~Bechstedt, L.~Reining,
\newblock \emph{Phys. Rev. B} \textbf{2012}, \emph{85}, 15 155131.

\bibitem{pavlyukh_interacting_2022}
Y.~Pavlyukh, E.~Perfetto, G.~Stefanucci,
\newblock \emph{Phys. Rev. B} \textbf{2022}, \emph{106}, 20 L201408.

\bibitem{yan_single-impurity_2022}
J.~Yan, V.~Jani\v{s},
\newblock \emph{Phys. Rev. B} \textbf{2022}, \emph{105}, 8 085122.

\bibitem{rohringer_local_2012}
G.~Rohringer, A.~Valli, A.~Toschi,
\newblock \emph{Phys. Rev. B} \textbf{2012}, \emph{86}, 12 125114.

\bibitem{bickers_conserving_1989}
N.~E. Bickers, D.~J. Scalapino,
\newblock \emph{Ann. Phys.} \textbf{1989}, \emph{193}, 1 206.

\bibitem{drchal_dynamical_2005}
V.~Drchal, V.~Jani\v{s}, J.~Kudrnovský, V.~S. Oudovenko, X.~Dai, K.~Haule,
  G.~Kotliar,
\newblock \emph{J. Phys. Condens. Matter} \textbf{2005}, \emph{17}, 1 61.

\bibitem{bickers_conserving_1991}
N.~E. Bickers, S.~R. White,
\newblock \emph{Phys. Rev. B} \textbf{1991}, \emph{43}, 10 8044.

\end{thebibliography}

%\textbf{References}\\

\end{document}